\documentclass{article}

\usepackage{arxiv}

\RequirePackage[T1]{fontenc}
\RequirePackage[utf8]{inputenc}
\RequirePackage{calc}
\RequirePackage{indentfirst}
\RequirePackage{fancyhdr}
\RequirePackage{graphicx,epstopdf}
\RequirePackage{lastpage}
\RequirePackage{ifthen}
\RequirePackage{float}
\RequirePackage{amsmath}
\RequirePackage{amssymb} 
\RequirePackage[right]{lineno}
\RequirePackage{setspace}
\RequirePackage{enumitem}
\RequirePackage{mathpazo}
\RequirePackage{booktabs} 
\RequirePackage{titlesec}
\RequirePackage{etoolbox} 
\RequirePackage{tabto} 
\RequirePackage{xcolor, colortbl} 
\RequirePackage{soul} 

\RequirePackage{multirow}
\RequirePackage{microtype} 
\RequirePackage{tikz} 
\RequirePackage{refcount} 
\RequirePackage{changepage} 
\RequirePackage{attrib} 
\RequirePackage{upgreek} 
\RequirePackage{array} 
\RequirePackage{tabularx}
\RequirePackage{pbox} 
\RequirePackage{ragged2e} 
\RequirePackage[subfigure]{tocloft} 
\RequirePackage{marginnote} 
\reversemarginpar 
\RequirePackage{marginfix} 
\RequirePackage{enotez} 
\RequirePackage{xstring} 
\RequirePackage[labelformat=simple]{subfig} 

\title{Turbulence Modeling; Uncertainty Quantification; Computational Fluid Dynamics, Deep Learning, Convolutional Neural Networks}

\author{
 Minghan Chu \\
 Department of Mechanical and Materials Engineering\\
 Queen's University\\
  \texttt{17mc93@queensu.ca} \\
   \And
 Weicheng Qian \\
 Department of Computer Science\\
 University of Saskatchewan\\
  \texttt{weicheng.qian@usask.ca} \\
}

\begin{document}
\maketitle
\begin{abstract}
Turbulence Models represent the workhorse for simulations used in engineering design and analysis. Despite their low computational cost and robustness, these models suffer from substantial predictive uncertainty, most of which is epistemic. At present, the Eigenspace Perturbation Method (EPM) is the only approach to estimate these turbulence model uncertainties, using physics based perturbation to the predicted Reynolds stresses. While the EPM address the question of how to perturb the Reynolds stresses for uncertainty estimation, it does not address how much to perturb. This shortcoming leads to very generous uncertainty bounds that result in sub-optimal designs. In this investigation, we use Convolutional Neural Networks (CNN) to predict the discrepancy between predicted and actual turbulent flows. These can be utilized to modulate the degree of the perturbations in the EPM leading to a Physics Constrained Deep Learning approach for Reynolds Averaged Navier Stokes model uncertainty quantification. We test this approach on turbulent flows over aero-foils and periodic hills to show the efficacy of our approach. 
\end{abstract}


\section{Introduction}
The flow of fluids represents a key problem in engineering design and scientific analyses, across fields such as Chemical Engineering, Mechanical Engineering, Aerospace Engineering, Biomedical Engineering, Astrophysics, etc. Most of the flows of engineering and scientific interest are turbulent, and are characterized by stochastic fluctuations of the velocity and pressure fields, a higher rate of material mixing and momentum diffusion, a wide spectrum of coupled scales of motion, etc. Even after decades of research, there is no analytical theory of turbulence and closed form solutions exist only for extremely simple turbulent flows. For complex real-life turbulent flows that are of interest to engineering and science, we rely on turbulence models for Computational Fluid Dynamics (CFD) simulations. Turbulence models are simplified constitutive relations that approximate quantities of interest using available measurements conditioned upon model closure parameters. Approaches to account for turbulence in engineering simulations include:
\begin{enumerate}
\item Direct Numerical Simulations (DNS) \cite{moin1998direct, moser1999direct, lee2015direct}: Turbulent flows involve a wide spectrum of scales of motion, ranging from large energy-containing eddies to small, dissipative eddies. Direct Numerical Simulations explicitly compute all scales of motion and are potentially very accurate, but also computationally expensive. Due to this expense, they are feasible for simple geometries and very small durations to be simulated, but not for real life engineering design that involves hundreds annd even thousands of simulations for a single design.

\item Large Eddy Simulations (LES) \cite{lesieur1996new, lesieur2005large}: In LES, we explicitly compute the larger energy containing scales, but model the smaller dissipative scales of motion. While less expensive than DNS, LES is still computationally expensive, especially for large-scale simulations or simulations spanning long durations. The computational cost of LES increases with the desired grid resolution and the complexity of the flow. This limits the applicability of LES for most engineering design problems.

\item Reynolds Stress Models (RSM) \cite{speziale1991analytical, launder1975progress, mishra2017toward, johansson1994modelling, speziale1991modelling}: RSM approach models all scales of motion, but uses the Reynolds Stress Transport equation for the evolution of quantities of interest. While potentially both accurate and computationally inexpensive, these are limited to limited engineering studies because of their unreliability.

\item Eddy Viscosity Models (EVM) \cite{craft1996development, gatski2002linear, shih1995new, kraichnan1976eddy}: Eddy viscosity models model all sales of motion using the eddy viscosity hypothesis, that relates the instantaneous Reynolds Stress tensor to the instantaneous mean rate of strain tensor. These models are computationally inexpensive, robust and have been the workhorse of engineering design simulations for decades.  

\end{enumerate}

The focus of our investigation in on the limitations and uncertainties in Eddy Viscosity Models. These include extremely widely used turbulence models like the $k-\epsilon$ model \cite{launder1983numerical}, the $k-\omega$ model \cite{wilcox2008formulation}, the SST variant of the the $k-\omega$ model, etc. Due to the similarity, this analysis extends to sub-grid scale models used in Large Eddy Simulations also. The uncertainties in EVMs are both epistemic and aleatoric\cite{smith2013uncertainty}. Epistemic uncertainties are as we have incomplete knowledge about turbulence physics \cite{duraisamy2019turbulence, duraisamy2017status}, and have to make simplifications for engineering utility \cite{oliver2011bayesian, mishra2016sensitivity}, for computational tractability \cite{alonso2017scalable, kato2013approach}. A major source of epistemic uncertainty in EVMs is due to the mathematical form of the closure expression \cite{dow2011quantification}, where the model assumes that complex non-local turbulence physics can be captured by polynomial expansions of local tensors. This leads to the so-called model-form (or structural) uncertainty in turbulence models.  

The impact of such uncertainties in engineering design can be substantial. Engineering design is an iterative process, where an initial design is proposed, turbulence model based CFD simulations are carried out on this current design, based on the output of these simulations small changes are made to the design, etc. This cycle of CFD-simulation and small changes to the current design is run for hundreds of iterations till the design thresholds are satisfied. Errors and uncertainties in the turbulence model can accumulate over each such iteration, leading to suboptimal final designs. The cascading effect of the turbulence model uncertainty can even lead to misleading final designs \cite{wood1990modeling, du2000methodology, gurnani2005robust, mattia2007comparative, kokkolaras2006impact}. Design Under Uncertainty is an approach that aims to produce robust and reliable final designs in spite of these uncertainties \cite{yao2011review, padula2006aerospace}. Even for design under uncertainty, the final output depends on the calibration of the uncertainty produced by the uncertainty estimation procedure. Thus, engineering design needs a turbulence model uncertainty estimation methodology that does not lead to overly optimistic or pessimistic uncertainty estimates, but uncertainty estimates that are properly calibrated.  

Currently, the only physics based methodology for quantifying turbulence model structural uncertainties is the Eigenspace Perturbation Method (EPM)\cite{iaccarino2017eigenspace}. The EPM uses the turbulence model's prediction of the Reynolds Stresses along with physics based perturbations to estimate the uncertainties in the predictions of the turbulence model. This methodology has been successfully applied to many different problems like aerospace design \cite{mishra2019uncertainty, mishra2017rans, mishra2019estimating, mishra2017uncertainty, thompson2019eigenvector, demir2023robust, cook2019optimization, mishra2020design, righi2023uncertainties, li2024adjoint, li2025application}, civil design \cite{gorle2019epistemic, huang2020nonuniform}, aircraft certification \cite{mukhopadhaya2020multi, nigam2021toolset}, etc. In spite of these successes, the EPM has fundamental limitations. The key limitation is the EPM's reliance on physics based precepts to guide the perturbations to the Reynolds Stress Tensor. These physics based rules can only demarcate the range of perturbations that are physically permissible for a turbulent flow, but do not contain any information about what ranges are probable for a turbulent flow. This leads to uncalibrated uncertainty estimates that are larger than required. The degree of the perturbations seeks to account for the difference between the turbulence model simulations and the real flow. This difference is not uniform across the flow domain, but varies from region to region. Thus the degree of perturbations should vary over the flow domain too. This calls for a spatial functions that can predict the magnitude of perturbations \cite{gorle2019epistemic, chu2022model}. 

There are no physical precepts that can guide in the formulation of such a function that can modulate the magnitude of the eigenspace perturbations across  the domain of general turbulent flows. But from experience, most researchers have an intuitive understanding of the behavior of this functions. For instance the discrepancy between the turbulence model simulations and the real flow is expected to be larger near an inhomogeneity in the flow domain, such as a wall. Similarly the discrepancy between the turbulence model simulations and the real flow is expected to be larger in regions with high streamline curvature. In these regions, turbulence researchers would expect the magnitude of the eigenspace perturbations to be higher as well. This intuition arises from the researchers experience of analyzing hundreds of varied turbulent flow simulations and comparing them to the actual turbulent flow. Based on this observations, we believe that a Machine Learning (ML) model can learn this function to modulate the magnitude of the eigenspace perturbations across the domain of general turbulent flows. ML based models have been applied to turbulence modeling problems in the recent past with success \cite{duraisamy2019turbulence, brunton2020machine, chung2021data, chung2022interpretable, duraisamy2021perspectives, zhang2015machine}. A limitation of such ML models for Scientific domain applications is their lack of generalization. Scientific Machine Learning uses scientific domain knowledge to regularize and guide these models, so that their generalizability improves. This involves different approaches such as using soft or hard constraints for regularization \cite{baker2019workshop}, using the underlying differential equations like in Physics Informed Neural Networks (PINNs) \cite{raissi2019physics}, incorporating the underlying physics using differentiable codes \cite{hoidn2023physics}, etc. In our investigation, we present a Physics Constrained Deep Learning approach where the magnitude of eigenspace perturbations is predicted by a trained Convolutional Neural Network (CNN) and the method of perturbation application is via the EPM. Here the physics determines how to perturb the Reynolds stresses and ML models determine how much to perturb the Reynolds stresses at different points in the flow domain. We use a CNN model to correct the turbulence model's predictions of the turbulent kinetic energy. We use data to train a CNN and test its generalization across different flow cases, specifically turbulent flow over airfoils and past periodic hills. Both of these are benchmark problems for turbulence modeling.

\section{Method Details}

\subsection{Eigenspace Perturbation Method Details}
Turbulent flow are characterized by stochastic fluctuations of the velocity and pressure fields. Using Reynolds averaging \cite{pope2001turbulent}, the instantaneous velocity and pressure fields can be decomposed into mean and fluctuating components, $\Tilde{u} = U +u$, where $\Tilde{u}$ is the instantaneous velocity, $U$ is the mean velocity and $u$ is the fluctuating velocity field. The fluctuating velocity field has zero mean and thus zero expectation. The expectation of the covariance of the fluctuating velocity field is non-zero, and is referred to as the Reynolds Stress tensor, $R_{ij} = \left\langle u_i u_j\right\rangle$. This is a central quantity of interest and in incompressible turbulent flows, this is the quantity that turbulence models aim to predict. The Boussinesq hypothesis claims that the instantaneous Reynolds stresses are related to the instantaneous mean rate of strain tensor, linearly. 
\begin{equation} \label{Eq_uiuj}
    \left\langle u_i u_j\right\rangle=\frac{2}{3} k \delta_{i j}-2 v_{\mathrm{t}}\left\langle S_{i j}\right\rangle = 2 k\left(\frac{\delta_{i j}}{3}+v_{i n} \hat{b}_{n l} v_{j l}\right),
\end{equation}
where, the turbulent kinetic energy is $k$, the Kronecker delta tensor is$\delta_{i j}$, the coefficient eddy viscosity is $\nu_{t}$, and  $\left\langle S_{i j}\right\rangle$ is the mean rate of strain tensor. The symmetric positive definite tensor of the Reynolds stress can be decomposed into the eigenvectors $v_{i n}$ and eigenvalues $\hat{b}_{n l}$. 

The perturbation of the eigenvalues involves the use of the Barycentric triangle \cite{banerjee2007presentation} as a consistency constraint. Only states of turbulence anisotropy that lie within this Barycentric map are physically permissible or realizable. The eigen-directions of the anisotropy tensor are thus aligned with the eigen-directions of the mean rate of strain tensor. Consequently, $\delta_{i j} = -2\nu_{t} \left\langle S_{ij} \right\rangle$, $a_{i j} \equiv\left\langle u_i u_j\right\rangle-\frac{2}{3}$. Based on this simplification, ${a_{i j}}={2 k v_{i n} \hat{b}_{n l} v_{j l}}.$

Based on this analysis, we can see that the eigenvalues of the Reynolds Stresses, their eigenvectors and the turbulent kinetic energy affect the veracity of the Boussinesq hypothesis. The EPM addresses the limitations of this eddy viscosity hypothesis by perturbing the eigenvalues, eigenvectors and the amplitude of the turbulence model's predictions of the Reynolds Stress. Perturbations of the eigenvalues and eigenvectors leads to the form:

\begin{equation}
R_{ij} = 2\rho k  \left( v_{in}^* \Lambda_{nl}^* v_{lj}^* + \frac{1}{3} \delta_{ij} \right),
\end{equation}
where the starred quantities represent the perturbed quantities. The perturbations to $\Lambda$ are introduced with respect to the barycentric map $\textbf{x}$ with $\lambda_i^* = B^{-1} \textbf{x}^*$ where $B$ is the transformation from the eigenvalue space to the barycentric triangle. The perturbation aligns towards the three vertices of the Barycentric triangle $\textbf{x}_{1C}$, $\textbf{x}_{2C}$ and $\textbf{x}_{3C}$, representing the 1-, 2-, and 3-component states of turbulence anisotropy. The perturbed coordinates $\textbf{x}^*$ are  $\textbf{x}^* = \textbf{x} + \Delta_B (\textbf{x}^t-\textbf{x})$, where $\textbf{x}^t$ is the target vertex of the Barycentric triangle and $\textbf{x}$, the model prediction. In this expression $\Delta_B \in [0, 1]$ is the  perturbation magnitude.

\subsection{Convolutional Neural Network model details}
In this study, a one-dimensional convolutional neural network is used. The architecture of the model is selected using manual testing, and has  four layers and 86 parameters in total, all of which are arranged to train a predictive model that can learn from the paired Reynolds-Averaged Navier-Stokes (RANS) and Direct Numerical Simulations (DNS) simulations of turbulence kinetic energy, which are here represented as $k^{\mathrm{RANS}}$ and $k^{\mathrm{DNS}}$. The turbulent kinetic energy, $k^{\mathrm{RANS}}$, is the model's key input feature, and the corresponding value, $k^{\mathrm{DNS}}$, is the intended output during the training phase, laying the foundation for the model's learning. This trained CNN model is then used to evaluate the $k^{\mathrm{RANS}}$ data after the training. This allows the model to produce predictions that closely resemble and approximate the $k^{\mathrm{DNS}}$, demonstrating the model's generalizability. The model's robustness and generalization are demonstrated by its ability to accurately predict the turbulence kinetic energy across the remaining 80\% of locations, despite the fact that the training data was obtained from less than 20\% of the locations along the $x$-axis. In the context of turbulence modeling, this result highlights the possibility of using a smaller dataset for training while yet attaining a significant degree of predictive accuracy, thereby making a significant contribution to the area of CFD. Because they open the door to improvements in the comprehension and simulation of turbulent flows—which are crucial for a number of scientific and technical applications—the ramifications of this research go beyond simple predictive modeling.

\subsection{Turbulent Flow Test Cases Details}
We use two datasets for training, validation and testing of our methodology. These include datasets describing turbulent flows over an airfoil and past periodic hills. In both the cases, we have a paired dataset where we have high fidelity Direct Numerical Simulation predictions, as well as Reynolds Averaged Navier Stokes model predictions. at the same locations in the flow domain.

\begin{figure} 
\centerline{\includegraphics[width=4in]{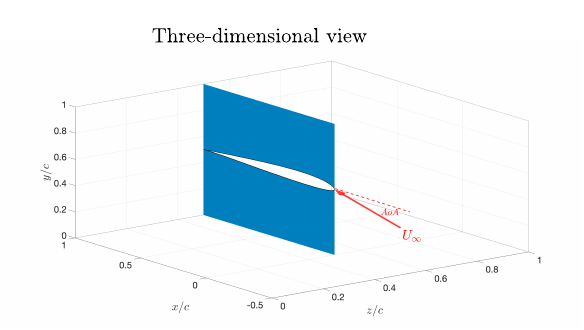}}
\caption{With freestream ($U_{\infty}$) coming into contact with the leading edge at $8^{\circ}$ AoA, a three-dimensional representation of the computational domain is shown.}
\label{Fig:SD7003_3DView}
\end{figure}

\subsubsection{Paired data for turbulent flow over a SD7003 Airfoil}

DNS was carried out simulating the turbulent flow over a SD7003 airfoil at an angle of attack of 8 degrees were used to generate this dataset (Figure ~\ref{Fig:SD7003_3DView}). The simulations were run using a Mach number of 0.2 and a Reynolds number based off of the chord length of $Re_c = 60000$. Information regarding the DNS dataset is available in \cite{zhang2021turbulent}. We ran comparative RANS simulations on the same conditions. Closure for the continuity and the momentum equations was implemented using transition model by Langtry and Menter \cite{langtry2006correlation, langtry2009correlation}. At an angle of attack of 8 degrees, the freestream inlet velocity ($U_{\infty}$) changed to turbulence on the airfoil's suction side, which corresponded to a Reynolds number of $Re_{c} = 60000$. A laminar separation bubble was seen on the airfoil's suction side when the conditions were close to stall. In spite of the use of the correlation-based transition model developed by Langtry and Menter \cite{langtry2006correlation, langtry2009correlation}, this flow provides a separation feature that is difficult to accurately depict for RANS models. The inability to capture separation and re-attachment accurately is a known limitation of eddy viscosity models. Despite being designed to describe different kinds of transitional flow, this model exacerbates the uncertainty in the Reynolds stress prediction \cite{chu2022model, chu2022quantification}. This uncertainty has a cascading effect on other quantities of interest such as the mean flow field and the pressure fields.

\subsubsection{Paired data for turbulent flow over Periodic Hills}
\begin{figure} 
\centerline{\includegraphics[width=6in]{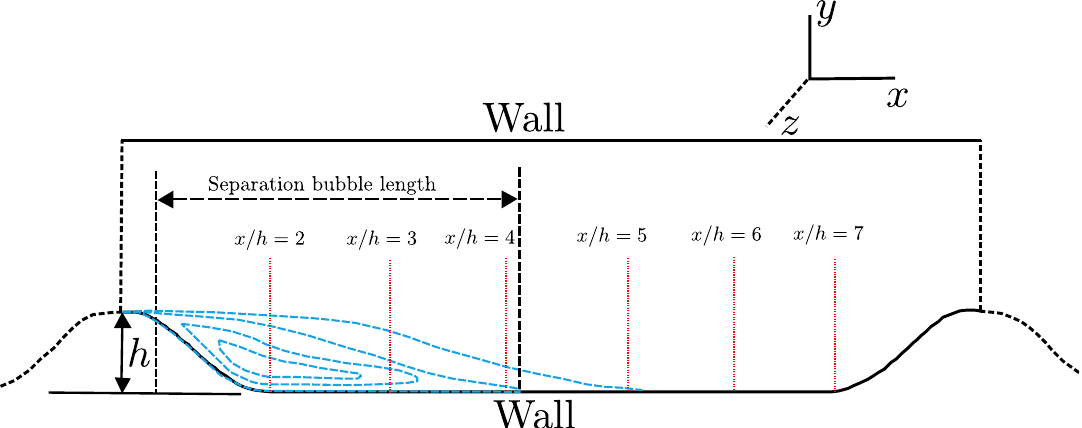}}
\caption{An illustration of the 2D periodic hills utilized in this study, with emphasis on the reference locations at $x/h = 2, 3, 4, 5, 6,$ and $7$. Streamlines show that a separation bubble is forming downstream of the slope. The hill height is $h = 0.028 \text{m}$.}
\label{fig:periodichill.pdf}
\end{figure}

Figure \ref{fig:periodichill.pdf} \cite{voet2021hybrid} reports on two open source datasets regarding flow dynamics across periodic hills, a benchmark case for turbulence modeling. These datasets come from two different sets of simulations: 1) 30 low-fidelity Reynolds-averaged Navier-Stokes (RANS) simulations, and 2) seven high-fidelity direct numerical simulations (DNS) \cite{voet2021hybrid}. The Reynolds number of $5600$ remains uniform across both simulation types. The simulations investigate changes in the turbulent flow with variation in two parameters describing the geometry: $\gamma$, the distance between neighboring hills, and $\alpha$, the steepness of the hills. While $\gamma$ establishes the reattachment point between successive hills, $\alpha$ controls the magnitude of the adverse pressure gradient that forms as the flow descends the hill. The convolutional neural network (CNN) model is trained in this study using a specific combination of $\alpha$ and $\gamma$ that is fixed at 1.0. The efficacy of the model is then evaluated for a range of $\alpha$ values, with $\gamma$ being maintained at 1.0 continuously. Please refer to \cite{voet2021hybrid} for detailed information on the open source datasets used.

\section{Numerical Investigation}

\subsection{Configurations} \label{sec:VoetData}

\begin{figure}[h!]
    \centering
    \includegraphics[width=\linewidth]{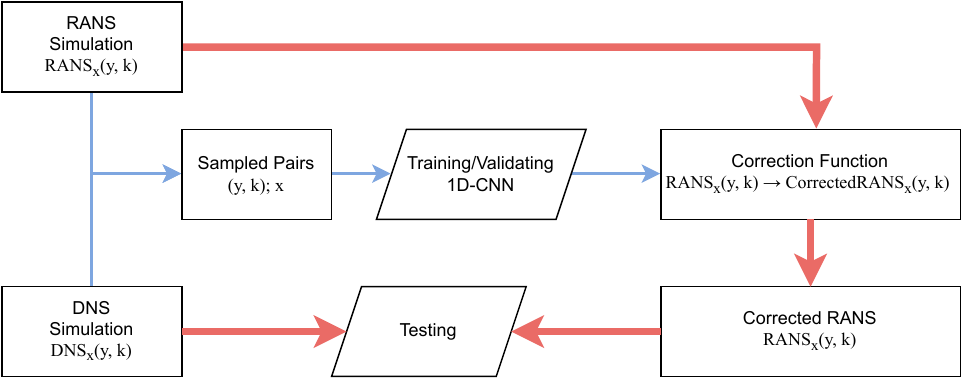}
    \caption{The methodology used in this study. The training paradigm is represented by the blue path, and the validation paradigm by the red path.} 
    \label{fig:data-flow.pdf}
\end{figure}

Following the simulation data production procedure shown in Figure \ref{fig:data-flow.pdf}, we perform tests using the CNN model. The CNN model is used to correct the RANS simulations predictions. The ground truth value, or label, for evaluating the precision of the CNN's corrected Reynolds-Averaged Navier-Stokes (RANS) predictions is the Direct Numerical Simulation (DNS) data. As a zeroth order baseline, the uncorrected RANS forecasts are used. We filter $x$-coordinates that are shared by both DNS and RANS meshes, since DNS and RANS usually run on different grid topologies. Based on the key $x$, the pairings of $(\mathbf{k}^{\mathrm{RANS}}, \mathbf{k}^{\mathrm{DNS}})$ are separated into a training set and a validation set, arranged by $x$-coordinate.  For training and validation, we use an 80\%–20\% split of the dataset, with 80\% of the data points from all $x$-locations going to a training dataset and the remaining 20\% going to a testing dataset.

The Mean Absolute Error (MAE) is used as the criterion (also known as the loss or the objective function) for generating gradients whilst model training, as well as a metric for the model performance. Both the uncorrected RANS and the RANS corrected by the CNN have their MAE loss computed and compared. 

For each direct numerical simulation (DNS) measurement, the CNN architecture in this study is particularly designed to accommodate input derived from RANS, using a window size of 11. The network's first layer is a one-dimensional convolutional layer with a kernel size of 3, a stride of 1, and ``valid'' padding, which means that the input data is not padded. High-level feature representations can be learned more efficiently using this setup. Two fully connected layers are used to further process the learnt feature maps after the convolutional layer. A max pooling layer is appended to reduce on computational cost. The Adam optimizer was used to train the model across 800 epochs with a batch size of 10 at a learning rate of 0.001. In order to strike a balance between model complexity and performance, extensive testing and subject-matter knowledge guided the selection of hyperparameters, which included kernel size, stride, padding, layer dimensions, and overall design. These hyperparameters were selected based on manual exploration, using a validation set. To improve forecast accuracy and processing efficiency, the design was tested and improved iteratively.

\section{Results}

\begin{figure} 
\centerline{\includegraphics[width=6in]{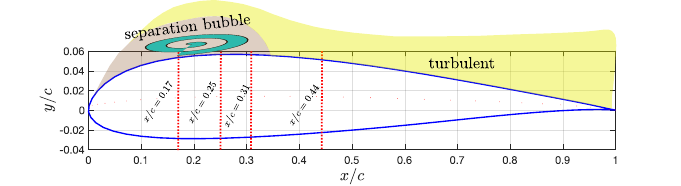}}
\caption{The flow over the SD7003 airfoil used in this study is shown, with $x/c = 0.17, 0.25, 0.31,$ and $0.44$ marked as locations that we focus upon, with separation points and regions of turbulent marked. The chord length is set at $C = 0.2 \text{m}$.}
\label{fig:SD7003_schematic.pdf}
\end{figure}

In this section, we analyze the turbulent flow over 2-D periodic hills and over the SD7003 airfoil. Our focus is on performance of the CNN model and its ability to predict the discrepancy between the baseline RANS predictions and the results of the corresponding DNS studies. 

Figure \ref{fig:rans-predicted-dns-main4-viz-loss.pdf} to Figure \ref{fig:corrected_dns_7} show the results for the test cases. Aside of the naive contrast between the corrected solutions using the CNN model and the RANS predictions, we focus on the $L^1$ norm at each $x/c$ location to enable a more nuanced evaluation of improvements using our ML augmented paradigm. By providing a precise quantitative measure of prediction accuracy, the $L^1$ norm helps identify areas for improvement that would not be immediately clear from direct comparisons alone. The limitations of eddy viscosity models are well known to the fluid dynamics community, such as near inhomogeneties, in regions of streamline curvature, and especially in predicting the location of the separation and reattachment points along with the length of the separation bubble. With this background, we have selected four specific positions for each of the two benchmark flows. Even though the turbulent flow over the SD7003 airfoil and periodic hills is relatively not complex, the eddy viscosity turbulence models exhibit trouble predicting the turbulent flow inside the separated region. This arises due to the adverse pressure gradients, the substantial streamline curvature and the vicinity of inhomogeneities. As shown in Figure \ref{fig:SD7003_schematic.pdf}, the four representative $x/c$ positions for the flow over the SD7003 airfoil are $0.17$, $0.25$, $0.31$, and $0.44$. Based on the prior investigation by \cite{galbraith2010implicit, garmann2013comparative, chu2022model}, these locations were chosen as they lie in the neighborhood of the separated region. The location at $x/c = 0.17$ is selected as this is the location of the separation point of the turbulent flow's separation. The point at $x/c = 0.25$ lies in the separation bubble, after the flow has separated. The point at $x/c = 0.31$ is the location of the re-attachment point of the flow. The point at $x/c = 0.44$ is located after the turbulent flow has re-attached and thus, lies in the fully turbulent boundary layer. This choice of the locations in the flow to focus on enables the analysis to provide details of the turbulence model's performance in different regimes of the turbulent flow. Using these, we can precisely quantify the improvements afforded by the CNN-corrected solutions in different regions around the flow separation.

Using modulations of the $\alpha$ parameter, we experimented in five different situations on the \texttt{Voet} dataset, that is focused on two-dimensional periodic hills benchmark turbulent flow case. As the flow moves over the hill, this parameter influences the adverse pressure gradient and thus, affects the separation point\cite{voet2021hybrid}. Choosing representative locations that preserve constant $x/h$ values across all cases was challenging due to the differences in grid resolution for each scenario. Four locations that fall within meaningful ranges were therefore identified. According to \cite{voet2021hybrid}, the reattachment point is estimated to occur at $x/h = 4$ for all scenarios, whereas full flow separation is expected around $x/h = 2$. With this in mind, we evaluated the CNN model for each scenario at these four locations that fall within the significant ranges listed below (for more information, see Figure \ref{fig:periodichill.pdf}). We highlight that in the range $3.0 < x/h < 5.8$, that represents the separation bubble, the turbulence model is extremely limited and inaccurate, as has been described in prior studies focusing on the effeccts of flow separation on the accuracy of turbulence models.

\begin{table}
\begin{center}
\caption{Demarcation of different test cases based on the $\alpha$ and $\gamma$ parameters for turbulent flow over periodic hills.}

\label{table:hillparameters}

\begin{tabular}{c c c c}
\hline
Case & $\alpha$ &$\gamma$ \\
\hline
One & $1.0$& $1.0$  \\
Two & $0.8$& $1.0$  \\
Five & $0.9$ & $1.0$ \\
Six & $1.1$ &$1.0$\\
Seven & $1.2$ &$1.0$\\

\hline
\end{tabular}

\end{center}
\end{table}


\subsection{Turbulent Flow Over SD7003 Airfoil}

\begin{figure}
    \centering
    \includegraphics[width=\linewidth]{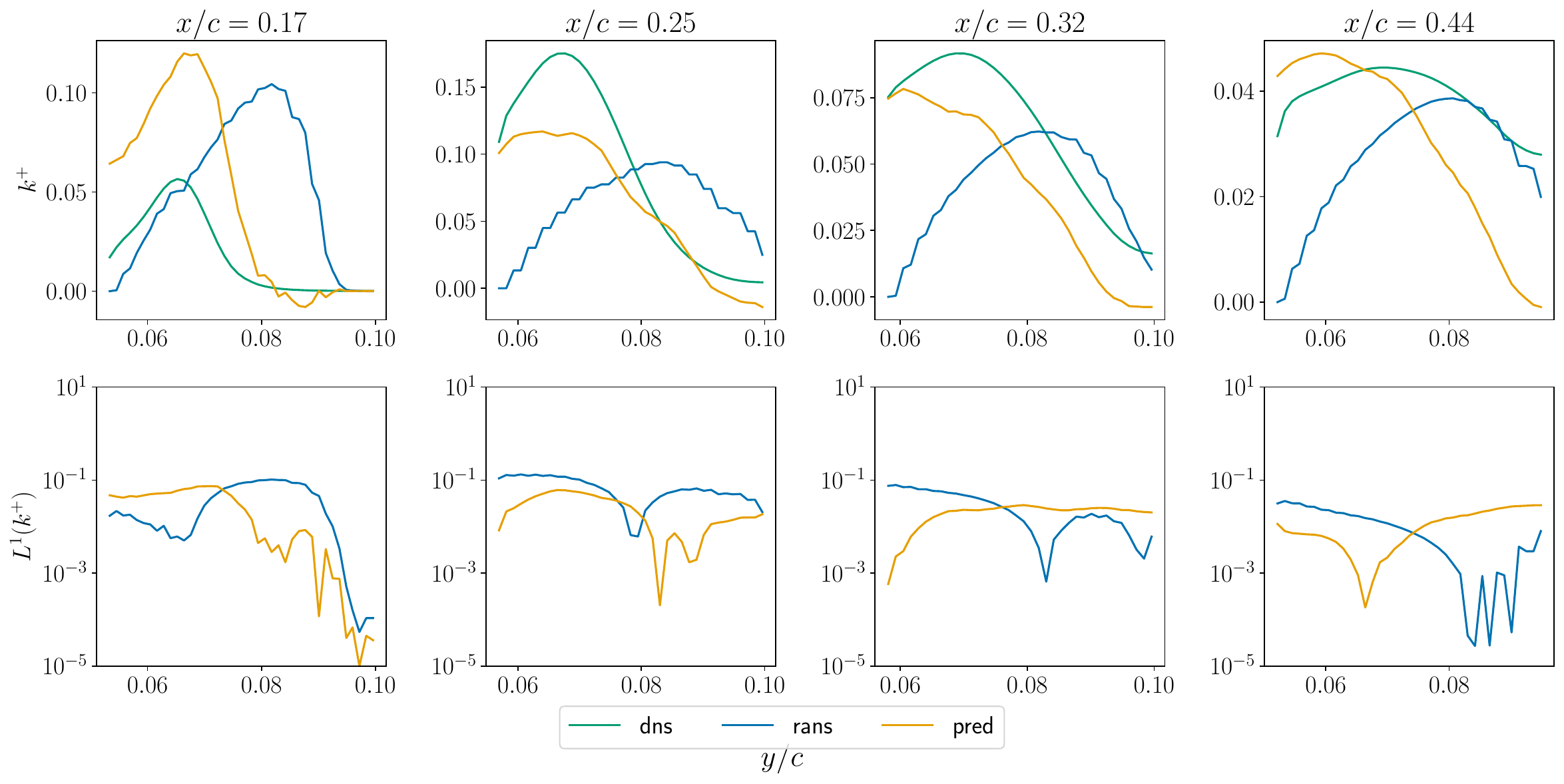}
    \caption{Results for the dataset \texttt{SD7003}. CNN forecasts for normalized turbulence kinetic energy based on DNS. The first row shows the CNN-predicted DNS (\texttt{pred}) and ground truth (\texttt{dns}), which have been smoothed using a moving average with a six-step window size. Second row: Comparing the $L^1$ loss between $L^1(\texttt{rans})$ and $L^1(\texttt{pred})$ validates the 1D-CNN model.}
    \label{fig:rans-predicted-dns-main4-viz-loss.pdf}
\end{figure}

We trained the CNN based corrective model using the data for the turbulent flow over an SD7003 airfoil  \cite{zhang2021turbulent,chu2022model}. The CNN-predicted turbulent kinetic energy profiles (normalized by the freestream velocity, represented as $k^{+}$), show a peak at the $y/c$ location that  closely agree with the ground truth DNS profiles across different $x/c$ positions in contrast to RANS, in Figure \ref{fig:rans-predicted-dns-main4-viz-loss.pdf}. The RANS results for $k^{+}$ show a peak in the turbulent kinetic energy that is shifted to the right by about $0.02$ of $x/c$ than the DNS ground truth data. The difference between the CNN-predicted $k^{+}$ profiles and the DNS ground truth improves with increasing $y/c$ at the transition point of $x/c = 0.17$. Additionally, this difference is much smaller than that between the RANS predictions and the DNS data. This highlights the improvements of the CNN corrections over the eddy viscosity models in these regimes of flow. The comparison of the RANS and CNN predictions' $L^1(k^{+})$ values supports this thesis, exhibiting that the CNN model's $L^1(\texttt{pred})$ is much lower, especially at the peak at $y/c \approx 0.07$ as well as beyond this point. The CNN-corrected prediction's $k^{+}$ profile consistently matches the ground truth DNS profile better than the RANS prediction beyond the region of the separation bubble. Thus, for the RANS model, the associated $L^1(\texttt{pred})$ is smaller than $L^1(\texttt{rans})$. With $L^1(\texttt{pred})$ being 2 orders of magnitude smaller than $L^1(\texttt{rans})$, the CNN model increases the accuracy of the predicted $k^{+}$ profiles near the wall at $x/c = 0.25$ within the bubble and $x/c = 0.44$ in the completely turbulent boundary layer farther downstream.

\subsection{Turbulent Flow Over Periodic Hills}

\begin{table}[h!]
\begin{center}
\footnotesize 
\setlength{\tabcolsep}{4pt} 
\renewcommand{\arraystretch}{0.9} 
\caption{Overview of the performance of the CNN model for adverse-pressure-gradient-induced separation turbulent flows.}
\label{table:summaryCNN}

\begin{tabular}{|c|c|}
\hline
\textbf{Flow cases} & \textbf{Observations}\\
\hline
SD7003 & 
\multicolumn{1}{m{10cm}|}{
    \begin{itemize}
        \item {The trained CNN model exhibits improved accuracy in predicting the location of maximal value of $k^{+}$ at the $y/c$  across $x/c$ locations.}
        \item {The trained CNN model has better accuracy in the turbulent flow's separation bubble and also the fully turbulent boundary layer. With the DNS as a baseline, the CNN model reduces errors by almost 2 orders of magnitude as compared to RANS-based predictions.}
    \end{itemize}
} \\
\hline
Periodic hills & 
\multicolumn{1}{m{10cm}|}{
    \begin{itemize}
        \item {The CNN model, trained on case 2, has better accuracy for cases with smaller alpha (e.g., cases 1, 2, and 5). This is highlighted in the turbulent flow separation bubble and in the vicinity of the reattachment point and near the wall.}
        \item {As alpha values increase (cases 6 and 7), showing a higher adverse pressure gradient, the model’s performance deteriorates.}
    \end{itemize}
} \\
\hline
\end{tabular}

\end{center}
\end{table}

To train the CNN-corrective model for this benchmark case of turbulent flow, we conducted an in-depth analysis using the RANS/DNS dataset \cite{voet2021hybrid}. As shown in Table \ref{table:hillparameters}, the study focuses on three examples that were taken from the datasets and defined by the parameters $\alpha$ and $\gamma$. Thirty RANS cases and seven DNS cases that are not immediately matched by their case indices make up the original datasets. For example, depending on their $\alpha$ and $\gamma$ values, the DNS dataset marked as scenario 2 is equivalent to RANS dataset marked case 6. We only use the DNS indices to represent the $\alpha$ and $\gamma$ parameters in order to avoid any confusion (see Table \ref{table:hillparameters}), which are also consistent with the corresponding RANS datasets that have the same $\alpha$ and $\gamma$ values. By altering only the $\alpha$ parameter in the hill geometry and keeping the $\gamma$ parameter fixed at one, we introduce uncertainty into the RANS datasets and assess how well our CNN model captures the turbulent kinetic energy of the turbulent flow. 

\begin{figure} 
    \centering
    \includegraphics[width=\linewidth]{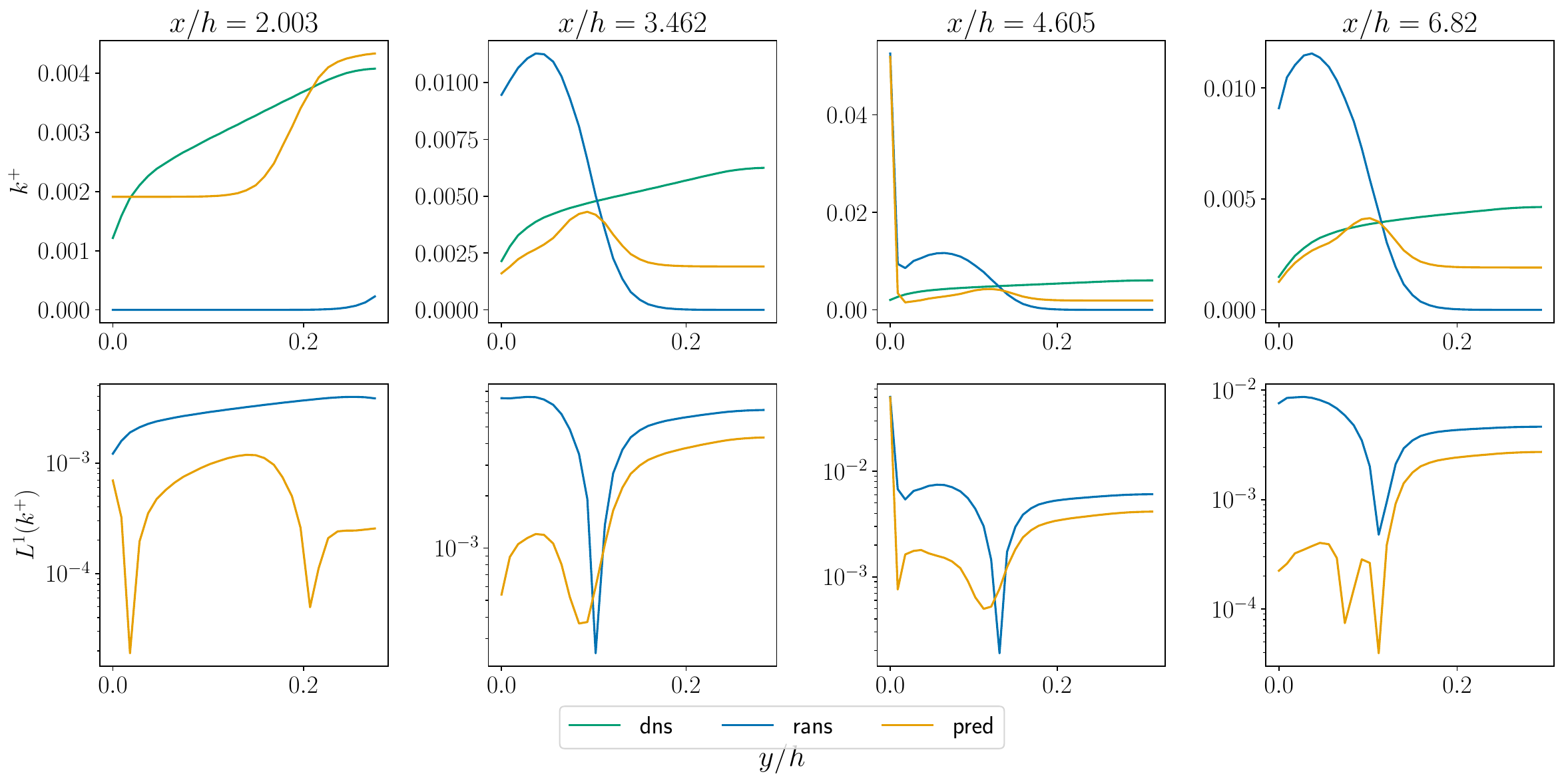}
    \caption{Tests for periodic hills in the \texttt{Voet} dataset. CNN prediction for normalized turbulence kinetic energy ($k^{+}$) based on case 2. First row: A moving average with a six-step window size was used to smooth the corrected DNS for Case 1 (\texttt{pred}) against the ground truth for Case 1 (\texttt{dns}). Second row: By comparing the $L^1$ loss between $L^1(\texttt{rans})$ and $L^1(\texttt{pred})$, the 1D-CNN is validated.}
    \label{fig:corrected_dns_1}
\end{figure}

\begin{figure} 
    \centering
    \includegraphics[width=\linewidth]{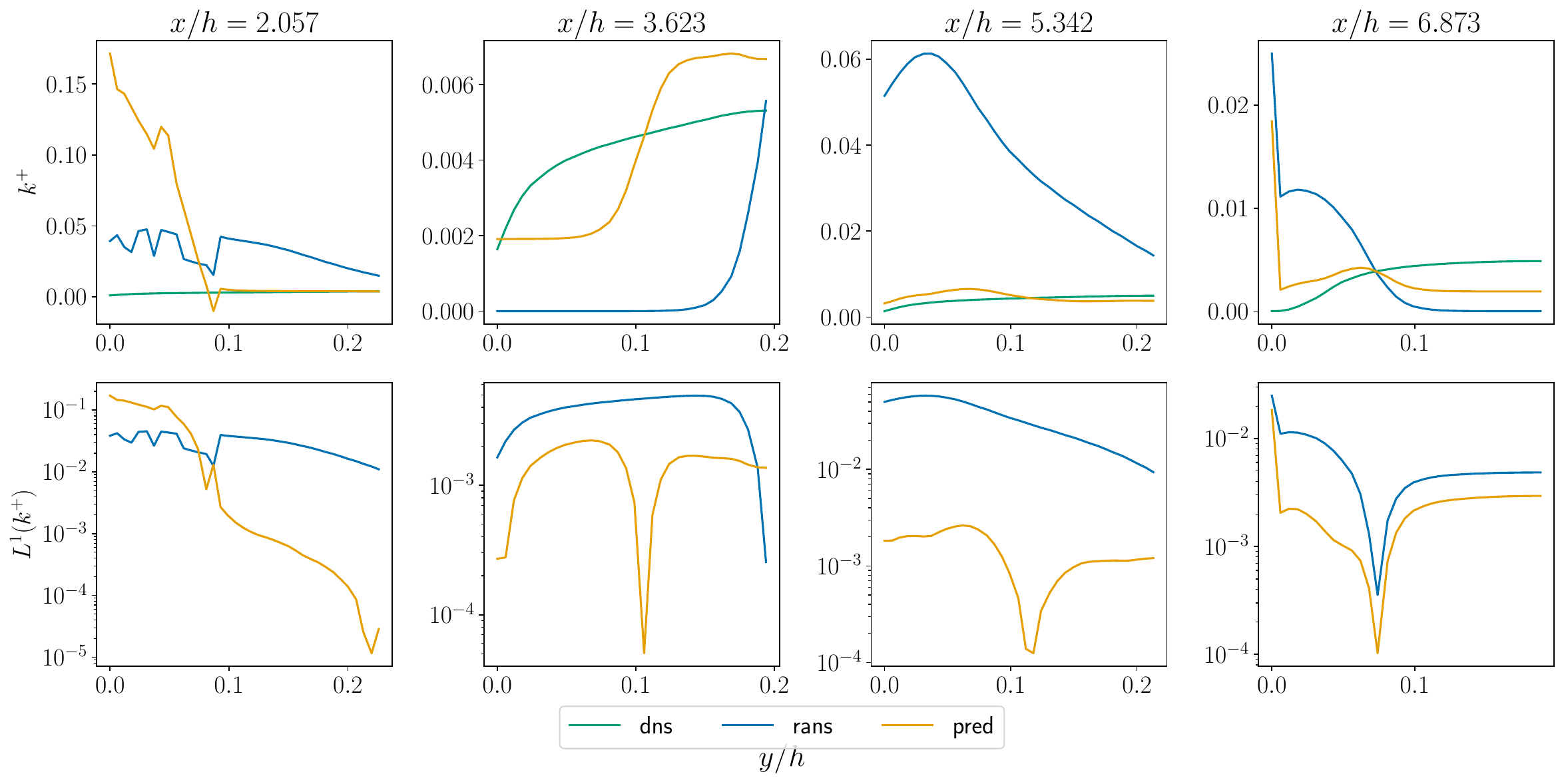}
    \caption{Observations for the case 2-based CNN prediction for normalized turbulence kinetic energy ($k^{+}$) for periodic hills in the \texttt{Voet} dataset. The first row is the corrected DNS case 2 (pred), which was smoothed using a moving average with a six-step window size and compared to the ground truth (dns) case 2. By comparing the L1 loss between $L^1(\texttt{rans})$ and $L^1(\texttt{pred})$, the second row validates 1D-CNN. }
    \label{fig:corrected_dns_2}
\end{figure}

\begin{figure} 
    \centering
    \includegraphics[width=\linewidth]{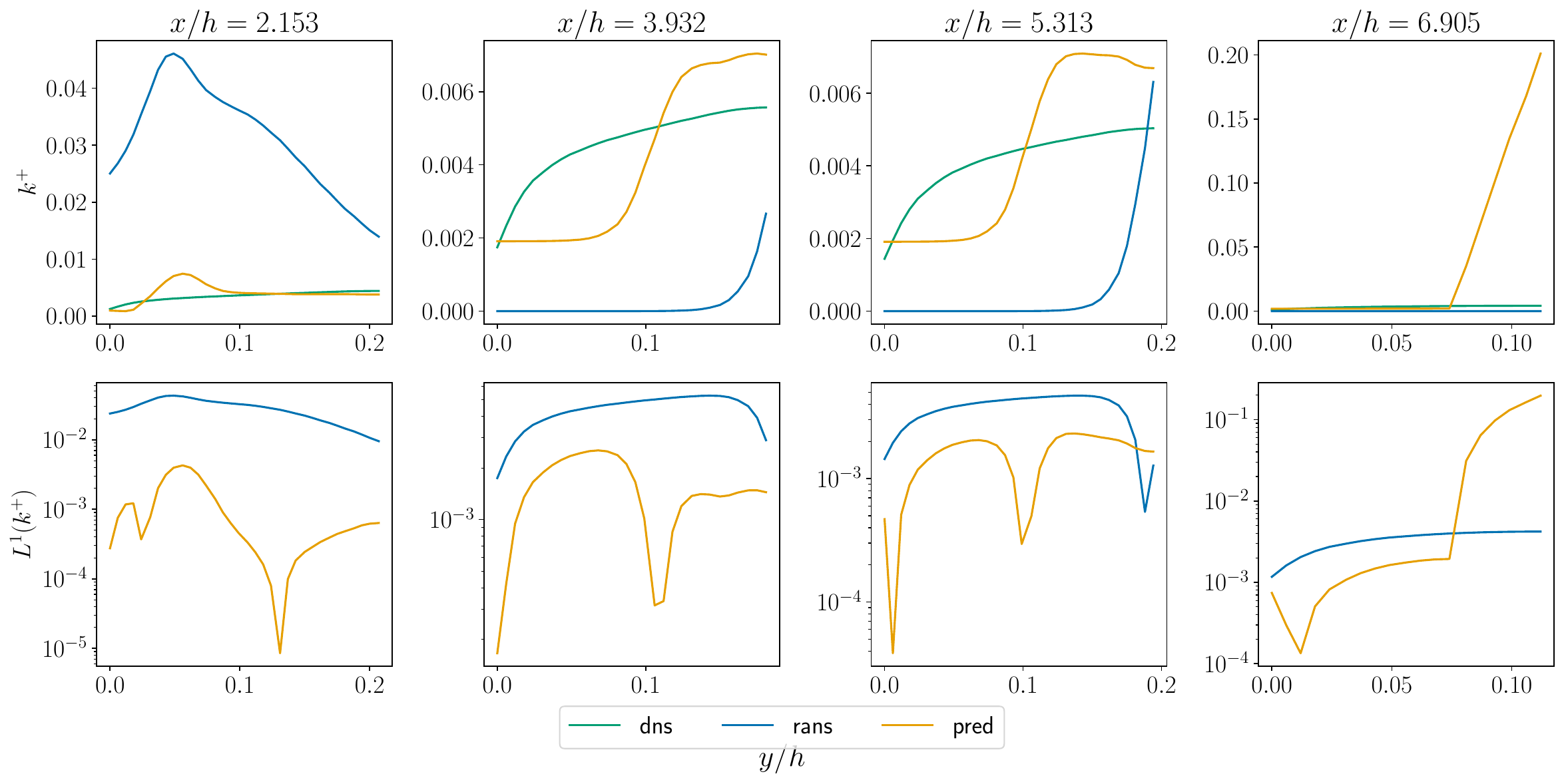}
    \caption{Results for the case2-based CNN prediction for normalized turbulence kinetic energy ($k^{+}$) for periodic hills in the \texttt{Voet} dataset. The first row is a comparison between the ground truth (dns) case 5 and the rectified DNS case 5 (pred), smoothed using a moving average with a six-step window size. By comparing the L1 loss between $L^1(\texttt{rans})$ and $L^1(\texttt{pred})$, the second row validates 1D-CNN.}
    \label{fig:corrected_dns_5}
\end{figure}

The CNN model that has been trained on case two from the RANS/DNS dataset \cite{voet2021hybrid}, is used to estimate the $k^{+}$ profiles of DNS for cases 1, 2, 5, 6, and 7 respectively, in Figures \ref{fig:corrected_dns_1} to \ref{fig:corrected_dns_5}. The tke profiles normalized by the bulk velocity through the channel, represented as $k^{+}$ for case one (where both the $\alpha$ and $\gamma$ parameters are set to $1.0$), show noticeably better predictions than the RANS-based predictions at four chosen positions, as shown in Figure \ref{fig:corrected_dns_1}. The CNN predictions' $L^1(\texttt{pred})$ metric is consistently two orders of magnitude lower than the predictions of the eddy viscosity model, especially near the wall. While $x/h = 4.065$ is close to the reattachment point, the positions at $x/h = 2.003$ and $x/h = 3.462$ are inside the separation bubble. These sites are extremely crucial to practitioners in the area since they present a significant amount of error in RANS model predictions that makes it difficult to generate appropriate RANS forecasts. Here, our CNN based correction model produces satisfactory results, with CNN-predicted $k^{+}$ values substantially closer to the correct DNS data than those obtained from the RANS predictions. The CNN-predicted $k^{+}$ profile and the DNS profile in the near-wall area for $y/h < 0.15$, which corresponds to the height of the bubble indicated by \cite{voet2021hybrid}, show a strong connection at $x/h = 3.462$ and $x/h = 4.605$. When the flow has fully reattached further downstream at $x/h = 6.82$, our CNN model still performs better than the RANS predictions for $k^{+}$. Particularly for $y/h < 0.15$, where the reduction of $L^1(\texttt{pred})$ approaches almost two orders of magnitude, the corresponding $L^1(\texttt{pred})$ value stays below that of $L^1(\texttt{rans})$.

In the second test case, the value of the parameter $\alpha$ is kept at $0.8$. The separation bubble may lengthen as a result of this lower $\alpha$ value compared to the prior case. According to \cite{voet2021hybrid}, the location of DNS reattachment is around $x/h = 4.769$ in this case and $x/h = 4.148$ in the first. When compared to the RANS predictions, Figure \ref{fig:corrected_dns_2} shows that the CNN model reduces the difference between the CNN-predicted $k^{+}$ profile and the actual DNS profile. In the separation bubble at $x/h = 2.057$, the CNN-estimated $k^{+}$ profile matches the DNS prediction in the outer boundary layer for $y/h > 0.1$, showing a significant improvement over the RANS prediction. Further downstream, the CNN model  improves the $k^{+}$ profile's predictive accuracy compared to the RANS-based prediction at $x/h = 3.623$ inside the separation bubble and around the reattachment point at $x/h = 5.342$. Compared to $L^1 (\texttt{rans})$, the corresponding $L^1 (\texttt{pred})$ values are 2 orders of magnitude lower. The CNN-predicted $k^{+}$ profile is generally closer to the actual DNS profile than the RANS-based prediction downstream to $x/h = 6.873$. This is more pronounced in the wall-adjacent region. For $y/h < 0.1$, the $L^1(\texttt{pred})$ is about one order of magnitude smaller than the $L^1(\texttt{rans})$.

In Figure \ref{fig:corrected_dns_5} we focus on the parameter $\alpha$ at 0.9. The difference between the predicted $k^{+}$ profile and the corresponding ground truth direct numerical simulation (DNS) profile at $x/h = 2.153$, located in the fully separated region, is significantly reduced by the CNN model. Compared to $L^1 (\texttt{rans})$, the calculated $L^1 (\texttt{pred})$ is 2 orders of magnitude smaller. When compared to the predictions made using the eddy viscosity model, the CNN's $k^{+}$ profiles are closer to the ground truth DNS profiles  inside the separation bubble, and close to the reattachment point at $x/h = 3.932$ and $x/h = 5.313$, as well as beyond the re-attachment point. The CNN model shows improvement in the near-wall region for $y/h < 0.08$ when the flow is reattached further downstream at $x/h = 6.905$. 

\begin{figure} 
    \centering
    \includegraphics[width=\linewidth]{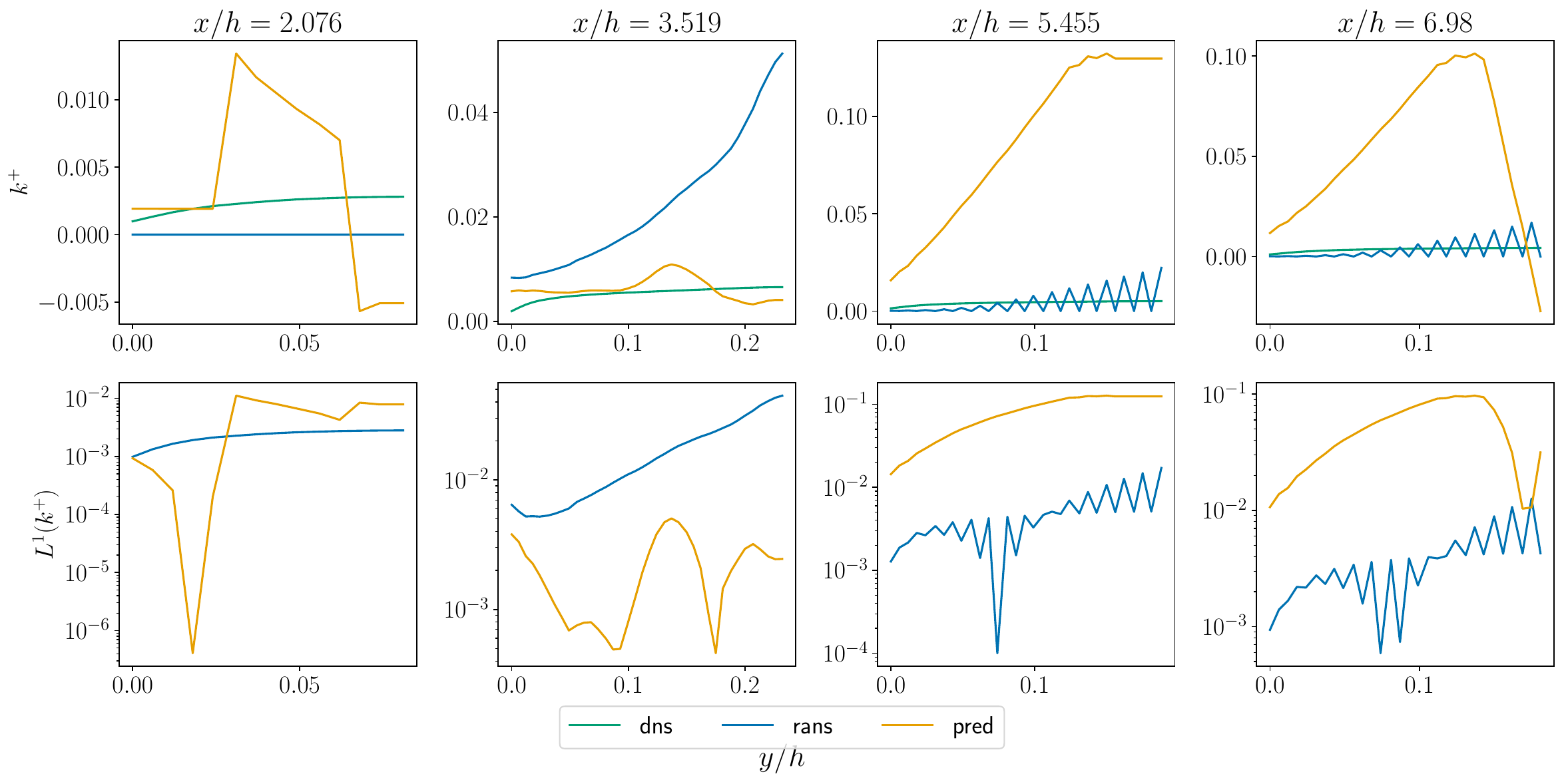}
    \caption{Outcomes for the case2-based CNN prediction for normalized turbulence kinetic energy ($k^{+}$) for periodic hills in the \texttt{Voet} dataset. The first row is a comparison between the ground truth (dns) case 6 and the rectified DNS case 6 (pred), which was smoothed using a moving average with a six-step window size. By comparing the L1 loss between $L^1(\texttt{rans})$ and $L^1(\texttt{pred})$, the second row validates 1D-CNN. }
    \label{fig:corrected_dns_6}
\end{figure}

\begin{figure} 
    \centering
    \includegraphics[width=\linewidth]{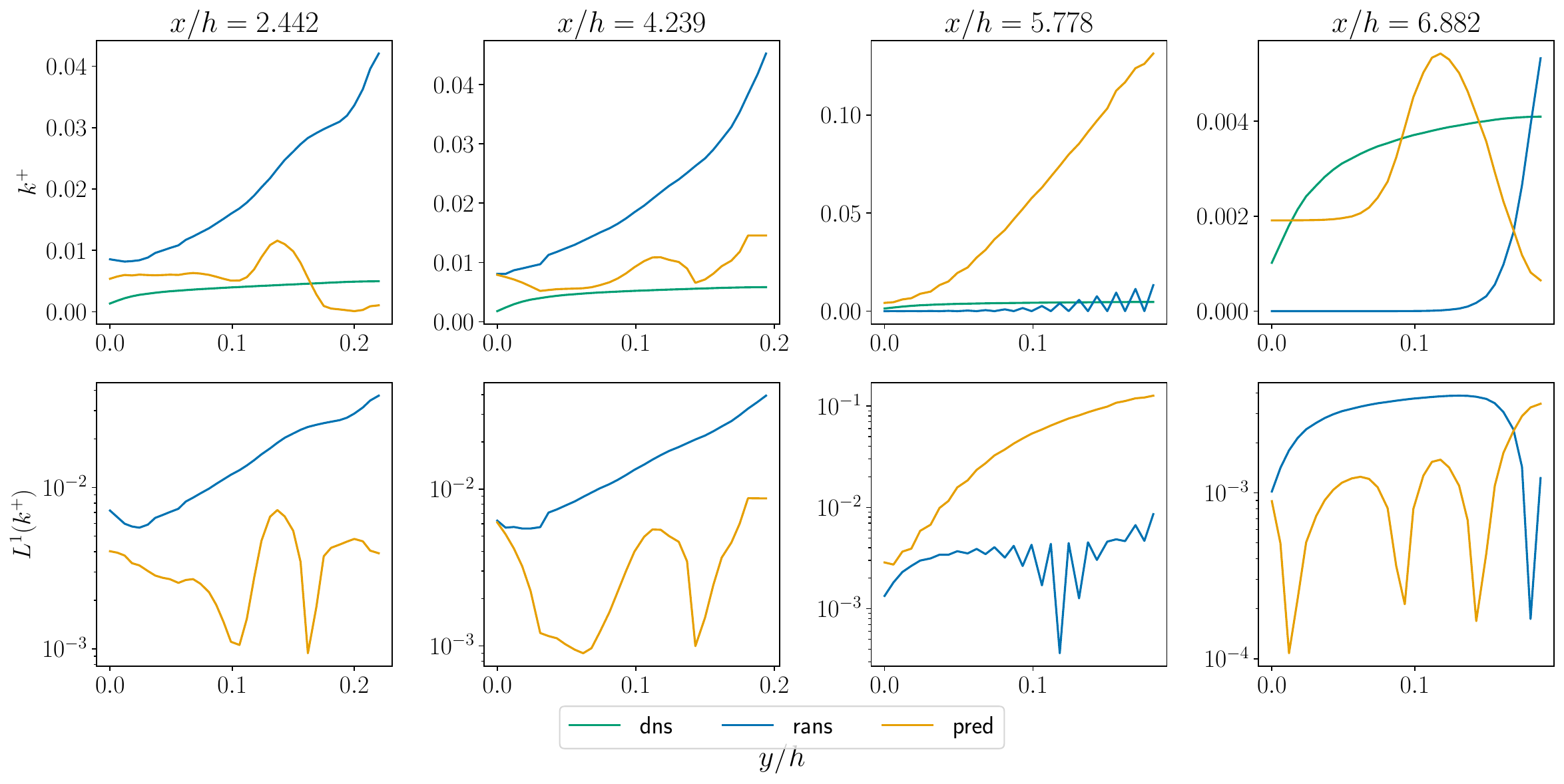}
    \caption{Results for the case2-based CNN prediction for normalized turbulence kinetic energy ($k^{+}$) for periodic hills in the \texttt{Voet} dataset. The first row is a comparison between the ground truth (dns) case 7 and the rectified DNS case 7 (pred), smoothed using a moving average with a six-step window size. By comparing the L1 loss between $L^1(\texttt{rans})$ and $L^1(\texttt{pred})$, the second row validates 1D-CNN. }
    \label{fig:corrected_dns_7}
\end{figure}

In the last test cases of this analysis, we increase $\alpha$ to 1.1 and 1.2, shown in Figures \ref{fig:corrected_dns_6} and \ref{fig:corrected_dns_7} respectively. Here, the CNN model's accuracy decreases in comparison to examples 1, 2, and 5. The largest difference between the $\alpha$ values for these cases and the training case 2 in these preliminary cases is 0.2, whereas the differences for examples 6 and 7 are 0.3 and 0.4, respectively. Based off of analysis of Figures \ref{fig:corrected_dns_6} and \ref{fig:corrected_dns_7}, the CNN model in the separation bubble shows improved prediction accuracy for case 6 at $x/h = 3.519$. The model's predictions are less accurate at other sites, where $y/h$ indicates notable departures from the ground truth DNS. At $x/h = 2.076$, $5.455$, and $6.98$, the corresponding $L^1(\texttt{pred})$ values are 1-2 orders of magnitude larger than $L^1(\texttt{rans})$. On the other hand, in case 7, the CNN model more accurately predicts the $k^{+}$ profiles in the separation bubble at $x/h = 2.442$ and $x/h = 4.239$, and downstream at $x/h = 6.882$, where the flow has completely reattached. In this case, CNN's projected $k^{+}$ profiles match the ground truth DNS profiles considerably better than RANS's predictions. Compared to $L^1(\texttt{rans})$, the corresponding $L^1(\texttt{pred})$ is roughly an order of magnitude smaller. However, close to the reattachment point at $x/h = 5.778$, the CNN-predicted $k^{+}$ profile deviates from the ground truth DNS profile with regard to $y/h$.

\subsection{Summary of Key Points for the Two Flow Cases}
The results of the 1D-CNN model applied to two different flow scenarios—the flow over an SD7003 airfoil and the flow over periodic hills—both of which are marked by the existence of a separation bubble, are summarized in Table \ref{table:summaryCNN}. The outcomes of our tests show that the 1D-CNN model exhibits a strong capacity for generalization in these diverse flow settings, which both show re-attachment and separation phenomena. Notably, as compared to conventional RANS techniques, the model consistently improves the predicted accuracy of turbulence kinetic energy, even though it was created using a very small dataset. The CNN model, which was trained on a particular flow example, successfully generalized to additional scenarios with comparable flow characteristics, especially those with lower $\alpha$ values, in the periodic hill scenario. It is crucial to remember that although the model's accuracy increased for these lower $\alpha$ cases, its performance declined for higher $\alpha$ scenarios, where the complexity resulting from stronger adverse pressure gradients produced less desirable results. Due to their inability to handle the non-local character of pressure-strain interactions, single-point RANS models commonly face difficulties in areas with separated flow. The CNN, on the other hand, makes use of convolutional kernels that detect spatial patterns among neighboring points, improving its capacity to simulate the interactions found in separated flows and producing predictions that are more accurate.

\section{Summary \& Conclusions}
The primary aim of this investigation is to use deep convolutional neural networks (CNNs) in the Eigenspace Perturbation Framework (EPM) to efficiently capture the spatially variable amplitude of the Reynolds stress tensor. This would enable modulation of the degree of the perturbations in the EPM, thus leading to more accurate uncertainty estimates. Previous studies have looked at how to improve the Eigenspace Perturbation Framework using machine learning models \cite{heyse2021data, heyse2021estimating, matha2023evaluation}. This investigation is the first attempt to use CNN models to explore the shift from RANS prediction space to DNS data space. We enable incorporation of non-local physics information into the uncertainty estimation methodology via the use of non-local convolutional kernels. This has been a longstanding limitation of eddy viscosity turbulence models, as these models are inherently single point closures and are unable to account for any non-local physics. The results show that the CNN models effectively learn the differences between RANS simulations and DNS data, which paves the way for the advancement of discrepancy marker functions in the future.

\bibliographystyle{unsrt}  
\bibliography{references}  






\end{document}